# Evaluating the performance of machine-learning-based phase pickers when applied to ocean bottom seismic data: Blanco oceanic transform fault as a case study

Min Liu[1] and Yen Joe Tan[1,*]

[1]Department of Earth and Environmental Sciences, The Chinese University of Hong Kong, Hong Kong S.A.R., China

*Corresponding author: yjtan@cuhk.edu.hk

**Key points**

- We systematically compare three machine-learning-based earthquake catalogs with an existing catalog based on traditional workflow
- The Pickblue-based catalog contains more earthquakes and/or provides better-constrained locations than the other catalogs
- Careful catalog assessment is crucial to prevent misinterpretation of fault slip modes from seismicity distribution

**Abstract**

Machine-learning-based phase pickers have been successfully leveraged to build high-resolution earthquake catalogs using seismic data on land. However, their performance when applied to ocean bottom seismic (OBS) data remains to be evaluated. In this study, we first adopt three machine-learning-based phase pickers – EQTransformer, Pickblue, and OBSTansformer – to build three earthquake catalogs for the 350-km-long Blanco oceanic transform fault (BTF) based on a year-long OBS deployment. We then systematically compare these catalogs with an existing catalog which utilized a traditional workflow. Results indicate that the Pickblue-based catalog documents more events and/or provides better-constrained locations than the other catalogs. The different performances of the three phase pickers suggest that detailed assessment of catalogs built using automatic workflows is necessary to prevent misinterpretations, especially when applied to regions without training samples. The Pickblue-based catalog reveals seismicity gaps in three extensional segments of BTF which likely represent aseismic slip zones affected by seawater infiltration. Furthermore, most earthquakes are shallower than the 600℃ isotherm predicted by a half-space conductive cooling model, except for the Blanco Ridge segment which has hosted 80% of the Mw > 6.0 earthquakes along BTF since 1976. These Blanco Ridge deep earthquake clusters can be explained by hydrothermal cooling or the serpentinization of mantle peridotite due to seawater infiltration along conduits created by the deeper ruptures of large earthquakes. Our analyses also demonstrate the importance of careful examination of automatically produced earthquake catalogs since mislocated events can lead to very different interpretations of fault slip modes from seismicity distribution.

**Plain Language Summary**

The performance of machine-learning-based phase pickers has been demonstrated for processing seismic data on land but remains unclear when applied to process ocean bottom seismic (OBS) data. Here, we systematically assess the performance of three machine-learning-based phase pickers (i.e., EQTransformer, Pickblue and OBSTansformer) in processing the OBS data deployed along the 350-km-long Blanco Oceanic Transform fault (BTF). We compare the corresponding earthquake catalogs with an existing catalog developed by a traditional workflow and find that the Pickblue-based catalog documents more earthquakes and/or provides better-constrained locations than the other three catalogs. The Pickblue-based catalog reveals some regions without/with minimal seismicity at shallow depths along BTF, which are likely affected by seawater infiltration. Earthquakes in the Blanco Ridge segment are significantly deeper than the estimated seismogenic depth. This can be explained by hydrothermal cooling or seawater infiltration along conduits created by the deeper ruptures of frequent large earthquakes in this segment. Therefore, both the different performances of the three phase pickers and the two alternative causes for deeper earthquakes in Blanco Ridge highlight the importance of careful examination of catalogs built using automatic machine-learning-based workflows to prevent misinterpretations from seismicity distribution.

## 1. Introduction

Earthquake detection and location are the cornerstones of seismological studies, such as the characterization of fault geometry (Liu et al., 2020; Ross et al., 2019; Tan et al., 2021), earthquake nucleation processes (Ellsworth & Bulut, 2018; Liu et al., 2022), and subsurface structure (body-wave tomography) (H. Zhang & Thurber, 2003). Traditionally earthquake detection was done by manually picking seismic arrivals from high-frequency filtered waveforms (e.g., Peng et al., 2006), which is relatively subjective and time-consuming. Later on, some automated methods, such as the short-term average and long-term average ratio (STA/LTA) (Allen, 1982; Withers et al., 1998) and match filtering technique (MFT) (Shelly et al., 2007), were developed and significantly enhanced the efficiency and performance of earthquake detection. However, these techniques also have certain limitations. For instance, STA/LTA cannot identify events with overlapped waveforms. While MFT is good at detecting overlapped events, prior template events are a prerequisite for performing MFT and MFT also has a high computational cost. More recently, machine-learning-based algorithms spawned a new generation of workflows to rapidly characterize earthquakes, such as LOC-FLOW (M. Zhang et al., 2022), QuakeFlow (Zhu, Hou, et al., 2022), and PALM (Zhou et al., 2022). These workflows usually incorporate a machine-learning-based phase picker (e.g., PhaseNet and EQTransformer) (Mousavi et al., 2020; Zhu & Beroza, 2018), a phase associator (e.g., REAL and GaMMA) (M. Zhang et al., 2019; Zhu, McBrearty, et al., 2022) and some classic earthquake location algorithms (e.g., VELEST, Hypoinverse, hypoDD and GrowClust) (Kissling et al., 1994; Klein, 2002; Trugman & Shearer, 2017; Waldhauser & Ellsworth, 2000). Current machine-learning-based phase pickers were mainly trained using seismic data on land and their superiority compared to routine catalog produced by earthquake monitoring agencies has been well-demonstrated, such as for the 2019 Ridgecrest and the 2016-2017 central Italy sequences (Liu et al., 2020; Tan et al., 2021). Although these machine-learning-based phase pickers have been applied to process ocean bottom seismic (OBS) data (H. Chen et al., 2022; Gong et al., 2023) and some new models for phase picking have been re-trained using OBS data (e.g., Pickblue and OBSTansformer) (Bornstein et al., 2024; Niksejel & Zhang, 2024), their performance when applied to process continuous OBS data has not been systematically evaluated yet.

Blanco transform fault (BTF) is a 350-km-long oceanic transform fault connecting the Juan de Fuca ridge and Gorda ridge in the northeast Pacific (Figure 1) (Embley & Wilson, 1992). BTF is composed of six segments including the West Blanco Depression (WBD), East Blanco Depression (EBD), Surveyor Depression (SD), Cascadia Depression (CD), Blanco Ridge (BR) and Gorda Depression (GD) (Figure 1), which exhibit various rupture behaviors (Braunmiller & Nábělek, 2008). A seismic network consisting of 55 OBSs was deployed along BTF from September 2012 to October 2013 (Figure 1) (Kuna et al., 2019). Based on this OBS network, Kuna (2020) built the first catalog (Kuna catalog) consisting of ~8,000 earthquakes for BTF by using a series of traditional methods, such as STA/LTA phase picking (Allen, 1982; Withers et al., 1998), Marquardt algorithm-based initial location (Pavlis et al., 2004) and hypoDD double-difference relocation (Waldhauser & Ellsworth, 2000). This catalog revealed swarm-like mantle earthquake sequences below BR, which may be driven by slow slip as inferred from the linear migration fronts during the evolutions of these sequences (Kuna et al., 2019). However, Kuna (2020) only used P-wave arrival times to locate the earthquakes, thus these locations can be improved further by incorporating S-wave arrival times. Based on the same OBS network, Ren et al. (2023) also constructed an earthquake catalog for BTF by combining the catalog provided by National Earthquake Information Center (NEIC) and a supplemental catalog developed by using the SeisComp3 package (GFZ & Gempa GmbH, 2008). However, this catalog only lists 144 events, which is significantly fewer than the Kuna catalog.

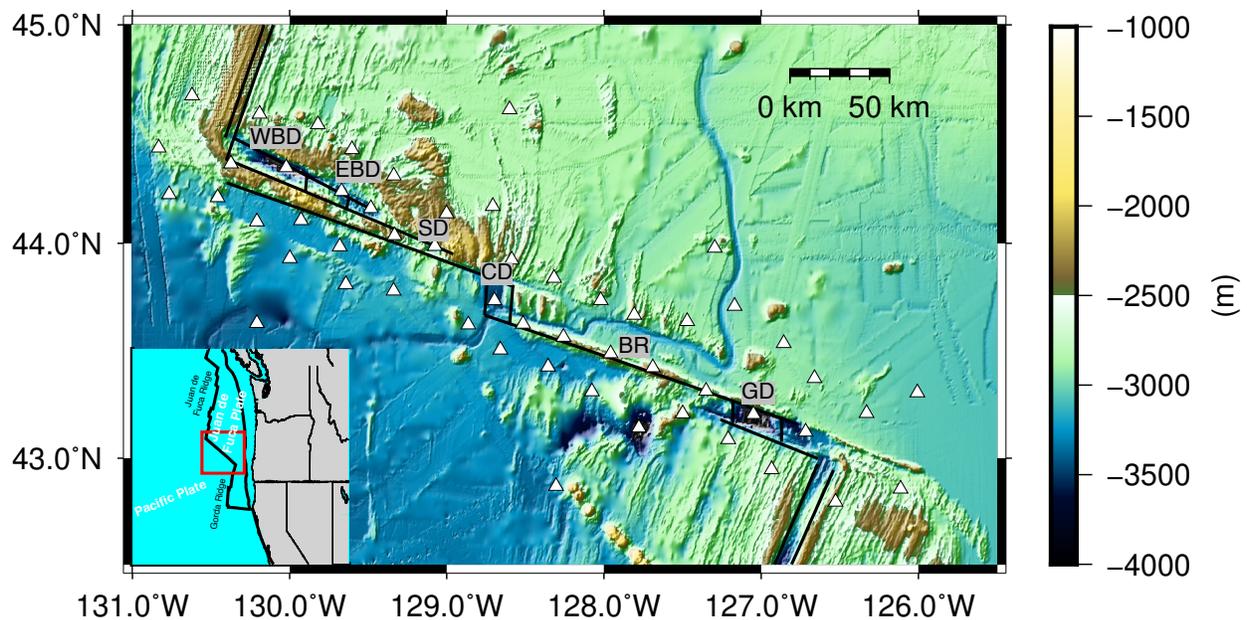

**Figure 1**. Map view of the study area. White triangles mark the OBS stations deployed during the Blanco experiment. Black lines outline the approximate locations of transform fault traces (Braunmiller & Nábělek, 2008), including the West Blanco Depression (WBD), East Blanco Depression (EBD), Surveyor Depression (SD), Cascadia Depression (CD), Blanco Ridge (BR) and Gorda Depression (GD). The inset shows the regional tectonic and our study area (red rectangle).

To assess the performance of machine-learning-based phase pickers when applied to continuous OBS data to develop a high-resolution catalog for BTF (Figure 1), this study first adopts three different machine-learning-based phase pickers (i.e., EQTransformer, Pickblue and OBSTansformer) (Bornstein et al., 2024; Mousavi et al., 2020; Niksejel & Zhang, 2024) to identify P- and S-wave arrival times. Subsequently, these arrival times picked by the three models are respectively inputted into a seismic data processing workflow consisting of phase association, initial location, and relative relocation, which results in three earthquake catalogs. We then compare the three machine-learning-based catalogs and the traditional workflow-based Kuna catalog (Kuna, 2020) to assess the performance of these machine-learning-based phase pickers. Finally, we analyze the relationship between the seismicity distribution, slip mode and the typical cooling model along BTF.

## 2. Data and Methods

The OBS network deployed along BTF (September 2012 ~ October 2013) consists of 30 broadband Güralp CMG3T seismometers and 25 short-period Mark Products L-28LB seismometers with a sampling rate of 100 Hz and an average inter-station spacing of ~25 km (Kuna, 2020). Note that station BB270 failed to record seismic data due to instrument issue. Previous studies have suggested that machine-learning-based phase pickers are more powerful in phase picking than traditional phase pickers (e.g., STA/LTA) (Liu et al., 2020; Tan et al., 2021). Here, we utilize three different machine-learning-based phase pickers (i.e., EQTransformer, Pickblue and OBSTansformer) (Bornstein et al., 2024; Mousavi et al., 2020; Niksejel & Zhang, 2024) to identify P- and S-wave arrival times from the continuous seismic data recorded by the 54 OBSs with available data. The three machine-learning-based phase pickers were trained using different datasets. EQTransformer was trained using the Stanford Earthquake Dataset (Mousavi et al., 2020), the first high-quality large-scale global data set of earthquake and non-earthquake signals recorded by seismic stations on land (Mousavi et al., 2019). Pickblue includes two transfer-learning models (i.e., EQTransformer and PhaseNet) (Mousavi et al., 2020; Zhu & Beroza, 2018) and its training dataset is compiled from manually picked P- and S-wave arrival times and non-earthquake signals on OBS data from 15 deployments in different tectonic settings worldwide (Bornstein et al., 2024). Note that the transfer-learning model of EQTransformer in Pickblue is adopted in this study. OBSTransformer is also a transfer-learning model of EQTransformer using an OBS-based training dataset (Niksejel & Zhang, 2024). Unlike Pickblue, however, the training dataset of OBSTransformer is compiled from the common phases picked by four widely used machine-learning-based phase pickers (EQTransformer, PhaseNet, Generalized Phase Detection and PickNet) (Mousavi et al., 2020; Ross et al., 2018; Wang et al., 2019; Zhu & Beroza, 2018) and the Akaike Information Criterion (AIC) picker (Maeda, 1985).

To better understand how different machine-learning-based phase pickers affect what is obtained in the final catalogs, the same parameters are used during phase picking as well as the later procedures. With a probability threshold of 0.1, EQTransformer identifies 973,651 P-wave arrival times and 953,878 S-wave arrival times. Both Pickblue and OBSTransformer identify many more phases than EQTransformer. Pickblue picks 2,852,054 P- and 4,710,099 S-wave arrival times, whereas OBSTransformer picks 5,294,374 P- and 9,430,016 S-wave arrival times. These P- and S-wave arrival times picked by the three models are further associated into individual earthquakes by using the Bayesian Gaussian Mixture model (GaMMA) (Zhu, McBrearty, et al., 2022). Within GaMMA, we conservatively set a threshold requiring at least 3 P phases, 2 S phases and 8 total phases. In total, GaMMA associates 23,538 (172,998 P phases and 184,910 S phases), 72,288 (499,653 P phases and 705,800 S phases) and 112,359 (687,596 P phases and 1,062,853 S phases) earthquakes from the picks identified by EQTransformer, Pickblue and OBSTransformer (Figure 2a), respectively.

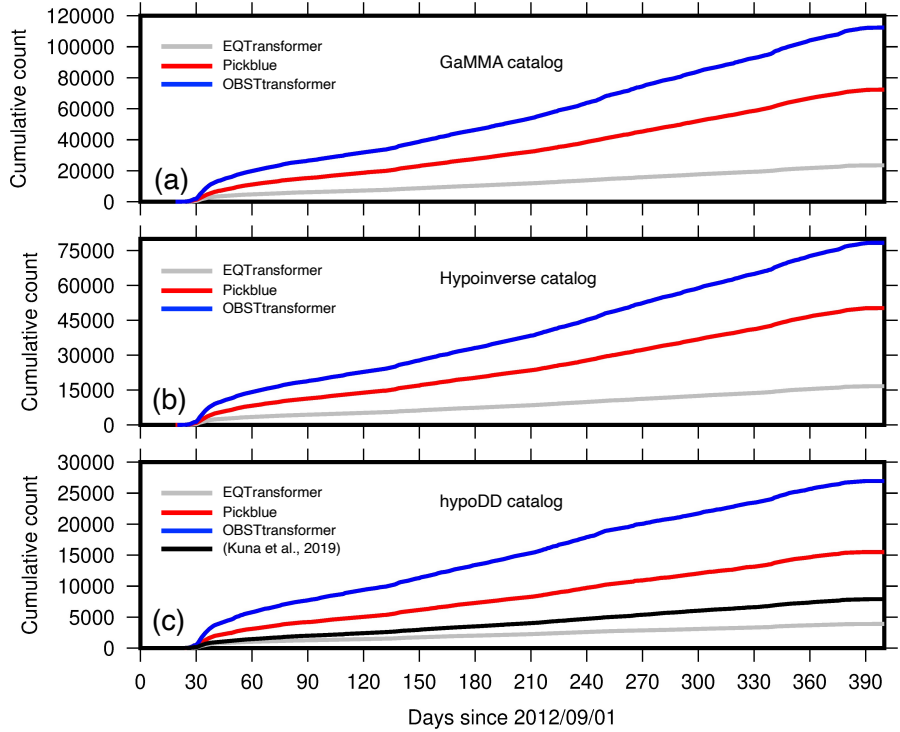

**Figure 2**. The cumulative count of earthquakes with time. (a) Gray, red and blue curves indicate the cumulative earthquakes with time for the EQTransformer-based, Pickblue-based and OBSTransformer-based GaMMA catalogs, respectively. (b-c) Similar to (a), but for hypoinverse catalogs and hypoDD catalogs. Black curve in (c) indicates the cumulative earthquakes with time for the Kuna catalog.

The associated earthquakes are only roughly located by GaMMA, thus we sequentially relocate these associated events by using an absolute location method – hypoinverse (Klein, 2002) and a relative location method – hypoDD (Waldhauser & Ellsworth, 2000). Only phases with epicenter distance < 150 km are used during the absolute location and the 1-D velocity model is the same as that used in (Kuna, 2020). After the absolute location, 16,683, 50,279 and 78,382 earthquakes with at least eight phases are kept in the EQTransformer-, Pickblue- and OBSTransformer-based catalogs (Figures 2b and S1), respectively. Before improving the relative locations of these events, we further select relatively well-constrained earthquakes with travel time residual < 0.6 s, station gap < 110° and phase number > 11. Besides, we only keep earthquakes with the nearest three stations having available phases, which allows better constraint of earthquake depths (Kuna, 2020).

In total, 4,021, 15,818 and 27,677 earthquakes from the EQTransformer-, Pickblue- and OBSTransformer-based catalogs are selected to be relatively relocated using hypoDD (Figure S2).

We then adopt the ph2dt algorithm provided by the hypoDD package to build event-pair database for the three catalogs (Waldhauser & Ellsworth, 2000). Within ph2dt, the maximum hypocenter separation and minimum links of event pairs allowed are 40 km and 12, respectively. The maximum number of neighbouring events for each earthquake is 20. The probabilities of the phases obtained from the three phase pickers are used as phase weighting in hypoDD. Finally, the hypoDD algorithm relocates 3,893, 15,505 and 26,944 earthquakes listed in the EQTransformer-, Pickblue- and OBSTransformer-based hypoinverse catalogs (Figures 2 and 3), respectively. The relative location errors of these events are estimated using a bootstrapping method (Waldhauser & Ellsworth, 2000). The average horizontal errors are 52 m, 88 m, and 287 m for the EQTransformer-, Pickblue- and OBSTransformer-based hypoDD catalogs, respectively. The average vertical errors are 62 m, 115 m, and 381 m for the EQTransformer-, Pickblue- and OBSTransformer-based hypoDD catalogs, respectively.

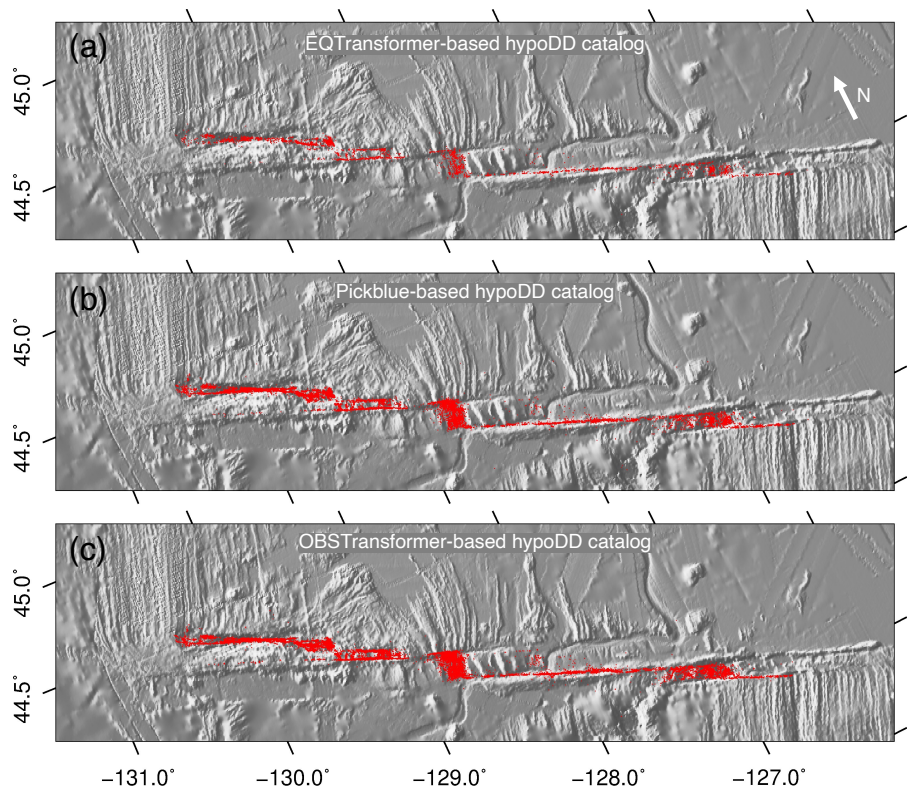

**Figure 3**. Earthquake location comparison between the EQtranformer-based, Pickblue-based and OBSTransformer-based hypoDD catalogs. (a) The epicenters of earthquakes listed in the EQtranformer-based hypoDD catalog. (b-c) Similar to (a), but for Pickblue-based and OBSTransformer-based hypoDD catalogs.

## 3. Results

Of the 1,927,529, 7,562,153 and 14,724,390 phases identified by EQTransformer, Pickblue and OBSTransformer respectively, 357,908 (~18.6%), 1,205,453 (~15.9%) and 1,750,499 (~11.9%) are further associated into 23,538, 72,288 and 112,359 earthquakes using GaMMA (Figure 2a). After the absolute location, 16,683 (~70.9%), 50,279 (~69.6%) and 78,382 (~69.8%) earthquakes are kept in the three hypoinverse catalogs respectively (Figure 2b). The epicenters of the three catalogs exhibit a similar pattern with the majority of earthquakes occurring on the NW-trending BTF, but the OBSTransformer-based hypoinverse catalog exhibits more off-fault seismicity (Figure S1). The on-fault earthquakes listed in the three catalogs outline a step-over fault system in map view (Figure S1). The hypocenters of the EQTransformer and Pickblue-based catalogs dominantly range from 0 to 10 km in depth (Figure S1), whereas the OBSTransformer-based catalog shows a deeper extension in depth with the deepest earthquakes reaching ~20 km (e.g., X= ~260 km; Figure S1c). The three catalogs also document the seismicity that occurred on the Juan de Fuca Ridge and Gorda Ridge, but the Pickblue-based hypoinverse catalog outlines a clearer geometry for the two ridges than the other two catalogs (Figure S1).

The locations of these relatively well-constrained earthquakes are further improved via hypoDD. Most off-fault and on-ridge seismicity are removed due to the requirement of station gap < 110°, but the different depth extensions of on-fault seismicity between the OBSTransformer-based catalog and the other two catalogs remain unchanged (Figure S2). After the relative relocation, the EQTransformer-, Pickblue- and OBSTransformer-based hypoDD catalogs finally document 3,893 (~23.3%), 15,505 (~30.8%) and 26,944 (~34.4%) earthquakes (Figure 2). The Pickblue- and OBSTransformer-based hypoDD catalogs include significantly more earthquakes than the two catalogs proposed by Kuna (2020) and Ren et al. (2023). While the EQTransformer-based hypoDD catalog documents more earthquakes than the catalog proposed by (Ren et al., 2023) as well, it is less than 50% of the Kuna catalog (Figure 2). These improved locations in the three hypoDD catalogs reveal a clearer and more linear geometry of BTF in map view than that revealed by their initial locations (Figures 3 and S2). The three hypoDD catalogs also indicate different extensions of seismicity in depth between the west (X < 170 km), middle (X = ~180 km) and east segments (X > 200 km). The western and eastern earthquakes dominantly range from 0 to ~8 km and from 0 to ~12 km in depth, respectively, whereas earthquakes within the middle segment extend down to ~20 km depth (Figure 3). In addition, the OBSTransformer-based hypoDD catalog documents a cluster of earthquakes extending down to ~20 km depth at X = ~260 km (Figure 3c), similar to the OBSTransformer-based hypoinverse catalog.

The EQTransformer-based hypoDD catalog documents much fewer earthquakes than the other two machine-learning-based hypoDD catalogs (Figure 2), suggesting that the performance of EQTransformer is worse than the other two models when applied to OBS datasets. To further assess the reliability of the two machine-learning-based catalogs documenting abundant earthquakes, we compare the relative locations along BTF between the Pickblue- and OBSTransformer-based hypoDD catalogs and the Kuna catalog. The three catalogs exhibit varying spatial patterns within different segments along BTF (Figure 4). The Pickblue-based hypoDD catalog reveals regions without/with minimal seismicity (seismicity gaps) within WBD, EBD and SD. These seismicity gaps extend from the seafloor and their widths vary along-strike with a maximum width of ~5.5 km (Figure 4a). The Kuna catalog seemingly reveals a similar pattern (Figure 4c). However, these seismicity gaps are not clear in the OBSTransformer-based hypoDD catalog except for the western SD (Figure 4b). Besides, the Pickblue-based hypoDD catalog includes some earthquakes deeper than ~8 km in the eastern WBD (Figure 4a), which are not observed in the other two catalogs (Figures 4b-c). Within CD, the earthquakes listed in the two machine-learning-based catalogs extend to ~20 km in depth (Figures 4a-b). While the Kuna catalog also has relatively deeper hypocenters in CD than to its west and east sides, earthquakes are generally shallower than ~15 km (Figure 4c). Finally, earthquakes in the Pickblue-based hypoDD catalog and the Kuna catalog show highly similar spatial patterns in BR and GD. The seismicity zone and gaps indicated by the two catalogs can basically match each other, including a ~100 km-long seismicity gap from 0 to

~5 km in depth between -127.7° and -126.8° in longitude (Figures 4a and c), but the Kuna catalog exhibits a larger seismicity gap between the crust and mantle earthquakes in BR (Figures 4a and c). In contrast, the OBSTransformer-based hypoDD catalog exhibit very different patterns from the other two catalogs, and the ~100 km-long seismicity gap is partially filled by earthquakes (Figure 4b). Besides, the OBSTransformer-based hypoDD catalog includes a cluster of earthquakes extending to ~20 km in depth around -127.9° in longitude, which is not observed in the other two catalogs (Figure 4).

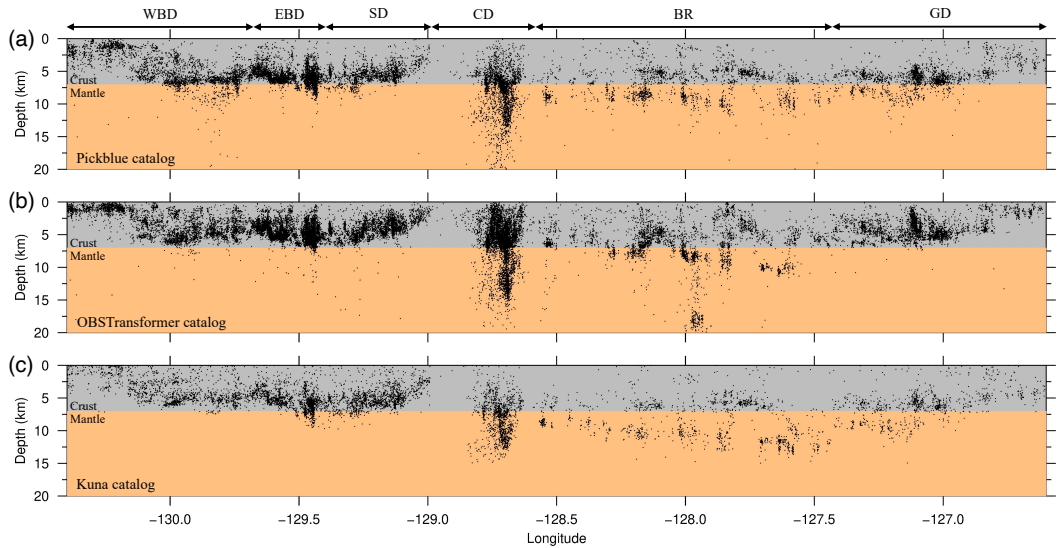

**Figure 4**. Earthquake hypocenter comparison along the west-east projection between Pickblue-based (a), OBSTransformer-based (b) and the Kuna (c) hypoDD catalogs. The West Blanco Depression (WBD), East Blanco Depression (EBD), Surveyor Depression (SD), Cascadia Depression (CD), Blanco Ridge (BR) and Gorda Depression (GD) are marked. Note that the y-axis is exaggerated two times in terms of distance compared to the x-axis.

## 4. Discussions

### 4.1 Assessing the reliability of machine-learning-based earthquake catalogs

Earthquake catalog is a fundamental part of many seismological studies (Ellsworth & Bulut, 2018; Liu et al., 2024; Z. Zhang et al., 2024) and recently machine-learning-based phase pickers have shown the promise to facilitate the rapid building of high-resolution earthquake catalogs using seismic data collected onshore (Liu et al., 2020; Tan et al., 2021). Although machine-learning-based phase pickers have also been applied to process OBS data (H. Chen et al., 2022; Gong et al., 2023), the reliability of the corresponding catalogs is unclear due to the lack of comparison with the traditional workflow-based catalogs. In this study, we adopt three machine-learning-based phase pickers (i.e., EQTransformer, Pickblue and OBSTransformer) with the latter two trained using OBS data to build three hypoDD catalogs for the ~350 km-long BTF (Bornstein et al., 2024; Mousavi et al., 2020; Niksejel & Zhang, 2024). To assess the reliability of the three catalogs, we systematically compare them with the Kuna catalog, which was established using a traditional workflow starting from STA/LTA (Kuna, 2020).

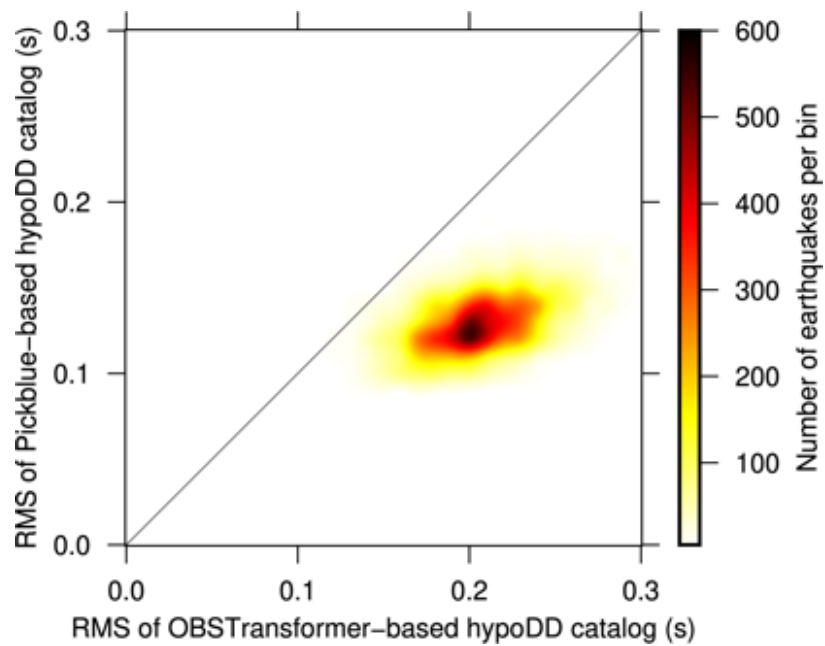

**Figure 5**. RMS comparison for common events between the OBSTransformer-based and Pickblue-based catalogs. The background color indicates the number of earthquakes per bin (0.01 s).

The EQTransformer-based hypoDD catalog only documents 3,893 earthquakes (Figure 2c), which is far fewer than the Pickblue- and OBSTransformer-based hypoDD catalogs (15,505 and 26,944), as well as the Kuna catalog (~8,000), suggesting that OBS training dataset is important to improve the performance of machine-learning-based phase pickers when applied to OBS datasets. While the OBSTransformer-based hypoDD catalog documents ~10,000 more events than the Pickblue-based hypoDD catalog, the former reveals a very different seismicity spatial pattern compared to the latter and the Kuna catalog (Figure 4). We compare the travel time residuals (i.e., RMS) of

12,716 common events between the Pickblue-based and OBSTransformer-based hypoDD catalogs. A common event pair is defined as two earthquakes occurring within a one-second time window between the two catalogs. In general, most events documented in the OBSTransformer-based hypoDD catalog have significantly larger RMSs than the Pickblue-based hypoDD catalog (Figure 5), and the mean increment is ~0.08 s. Besides, we also check the arrival times of these unusually deep earthquakes (longitude = ~127.9° and depth = ~20 km) in the OBSTransformer-based hypoDD catalog (Figure 4b), and find that both the P and S phases of these events at the nearest station BB200 are wrongly identified by OBSTransformer (Figure 6). Nearby seismic stations can provide a strong constraint for the determination of earthquake depth, hence the misidentified phases at the nearest station are likely the cause of the mislocated depths for these unusually deep earthquakes. Meanwhile, P and S phase misidentifications may also occur at other stations, resulting in the relatively large RMSs of events listed in the OBSTransformer-based hypoDD catalog (Figure 5), as well as the very different seismicity spatial pattern (e.g., the lack of shallow seismicity gap) compared to the Pickblue-based hypoDD catalog (Figure 4). Hence, the above comparison suggests that the Pickblue-based hypoDD catalog provides more reliable locations than the OBSTransformer-based hypoDD catalog.

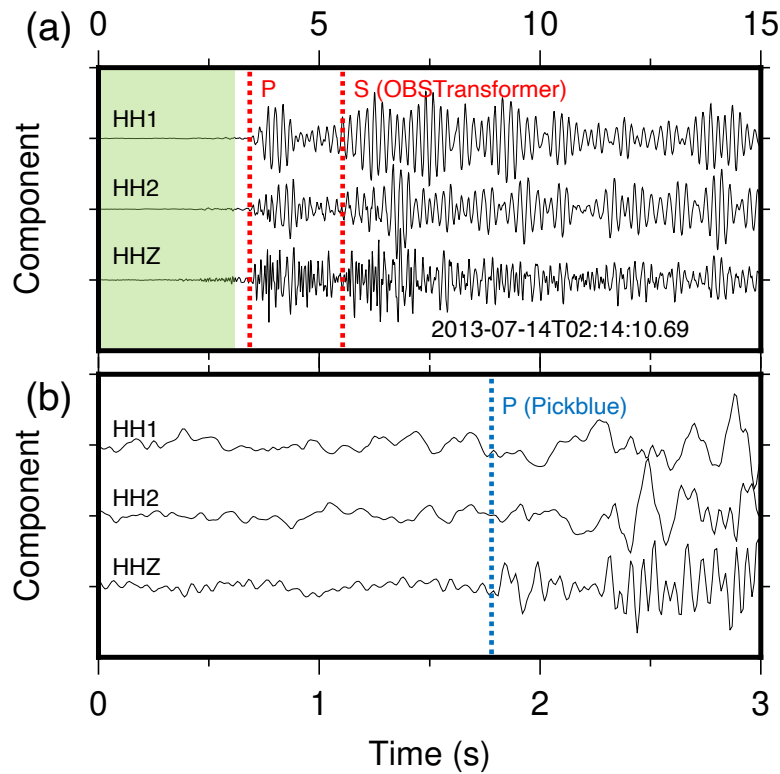

**Figure 6**. An example waveform and picks for one (2013-07-14T02:14:10.69) of those deep events below station BB200 (i.e., around -127.9° in longitude). (a) Three-component seismograms of an earthquake (2013-07-14T02:14:10.69) recorded by station BB200. Two red dashed lines mark the wrongly identified P and S phases by OBSTransformer. Green shadow indicates the time window exhibited in (b). Blue dashed line in (b) indicates the correct P phase of this event identified by Pickblue.

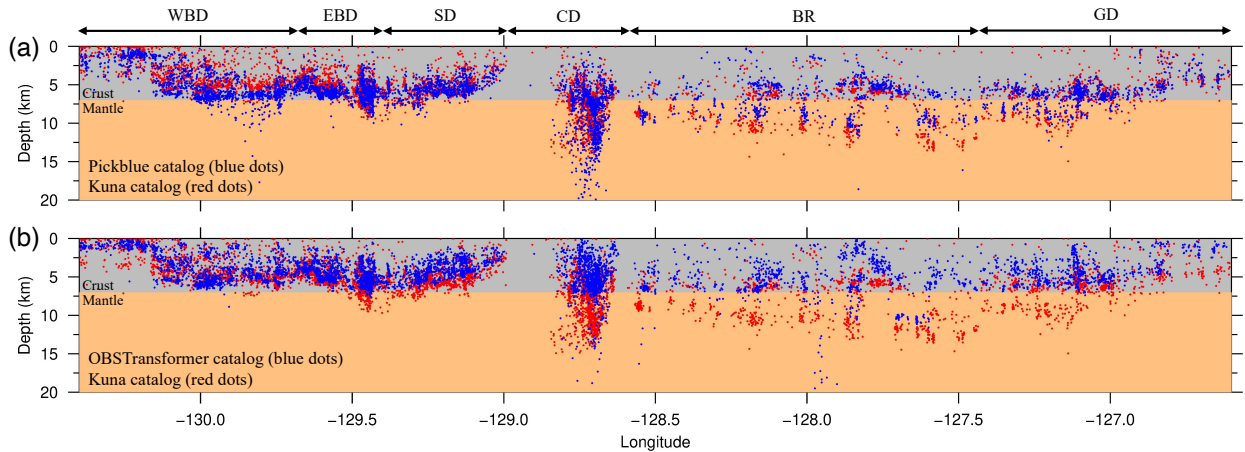

**Figure 7**. (a) Hypocenter comparison of common events between the Pickblue-based catalog (blue dots) and the Kuna catalog (red dots) along the east-west projection. The West Blanco Depression (WBD), East Blanco Depression (EBD), Surveyor Depression (SD), Cascadia Depression (CD), Blanco Ridge (BR) and Gorda Depression (GD) are marked. Note that the y-axis is exaggerated two times in terms of distance compared to the x-axis. (b) Similar to (a), but for the comparison between the OBSTransformer-based hypoDD catalog and the Kuna catalog.

We then compare the spatial distribution of 5,799 common events between the Pickblue-based hypoDD catalog and the Kuna catalog (Figure 7a). Earthquakes within the Kuna catalog that failed to match are mainly caused by the different criteria of seismic data processing between the two catalogs. In our workflow, only earthquakes with a total of 12 phases or more are selected for the final relative relocation, whereas the Kuna catalog kept earthquakes with 8 P phases (Kuna, 2020). By decreasing the threshold of phase number to 8 in our workflow, the successfully matched events between the Pickblue-based hypoDD catalog and the Kuna catalog can reach 7,104, which is ~90% of the Kuna catalog. The majority of the common events between the two catalogs outline a highly similar spatial distribution along the entire BTF but exhibit slight offsets in depth (e.g., Blanco Ridge; Figure 7a). Considering only P phases were used to build the Kuna catalog (Kuna, 2020), the small offset in depth of the Pickblue-based hypoDD catalog relative to the Kuna catalog may represent an improvement in locations due to the use of both P and S phases.

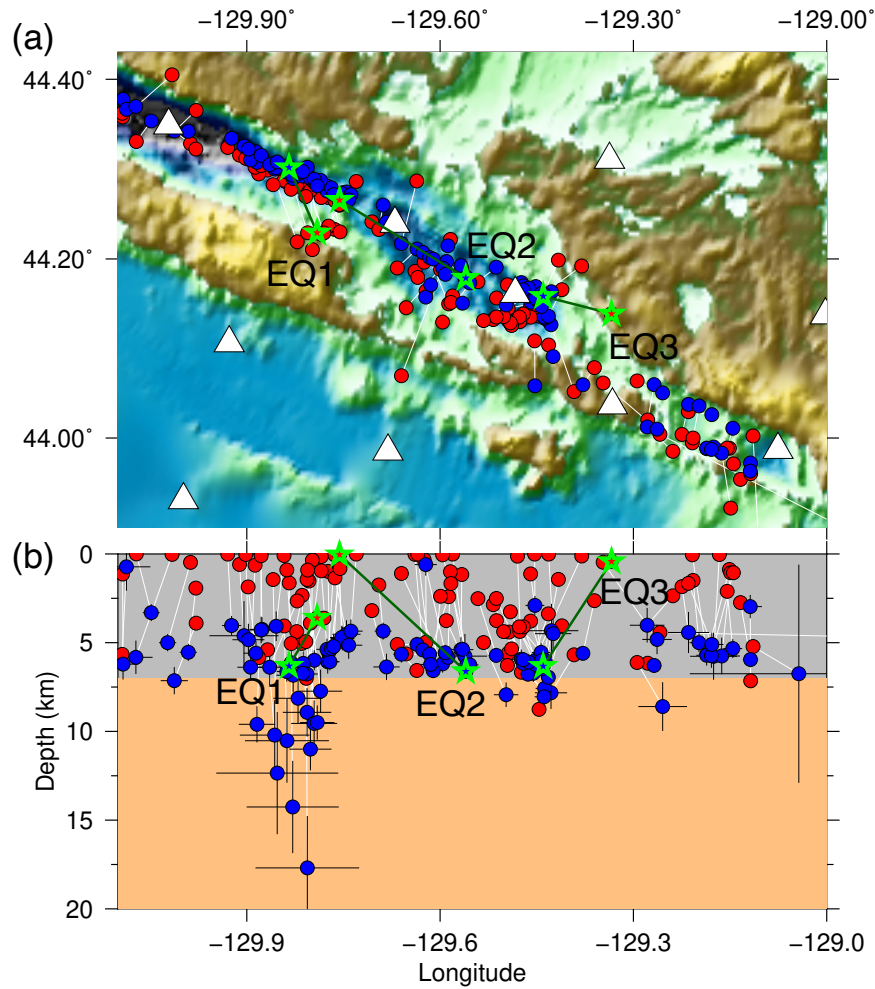

**Figure 8**. Comparison of the common events with relatively large separations between the Pickblue-based hypoDD catalog (blue dots) and the Kuna catalog (red dots) in the West Blanco Depression (WBD), East Blanco Depression (EBD), and Surveyor Depression (SD). (a) Map view of the common events, which are linked by white lines. White triangles mark the OBS stations deployed by the Blanco experiment. Six green stars linked by three dark green lines represent three common events (EQs 1-3), which are further examined by inspecting their distance-time curves in Figure 9. (b) Projection of these common events along the west-east direction. Black crosses indicate the relative errors of these events in the Pickblue-based hypoDD catalog.

A small portion of common events between the two catalogs also reveals different spatial distributions. The Pickblue-based hypoDD catalog contains some scattered events with significantly deeper depths than the Kuna catalog in the WBD (approximately -129.8° in longitude; Figure 7a) and CD (approximately -128.7° in longitude) whereas the Kuna catalog contains extremely shallow earthquakes (< 1 km) throughout the entire BTF (Figure 7a). To verify these different patterns revealed by the two catalogs, we first check the common events with relatively large hypocenter offset (i.e., > 4 km) in WBD, EBD and SD (Figure 8). These common events listed in the Pickblue-based hypoDD catalog outline a more linear geometry than the Kuna catalog in map view, whereas the latter documents more off-fault seismicity (Figure 8a). Within the Pickblue-based hypoDD catalog, we find that these unusually deep earthquakes (> 7 km) around the longitude of -129.8° exhibit significantly larger depth errors than shallower earthquakes (< 7 km; Figure 8b). It should be noted that the nearest station distant to these deep events is at least 10 km away (Figure 8a), implying that the depths of these deep events may be mislocated due to the lack of nearby station constraints. A similar issue is also observed in the Cascadia Depression, where earthquakes with depths > 15 km or deviating from the main earthquake cloud have significantly larger depth errors (Figure S3). Subsequently, we examine the off-fault seismicity and extremely shallow events in the Kuna catalog by inspecting the relations between arrival times and epicentral distances of three events (EQs 1-3) with relatively large horizontal separations in the two catalogs. For the Kuna catalog, EQ1 is from an off-fault cluster with a depth of ~3 km, EQ2 is an on-fault earthquake but with an extremely shallow depth of ~0.1 km, and EQ3 is an off-fault earthquake with an extremely shallow depth of ~0.4 km (Figure 8). For the Pickblue-based hypoDD catalog, however, all three earthquakes are on-fault events with a similar depth of ~6.5 km (Figure 8). Results indicate that the locations of the three events provided by the Kuna catalog do not match a reasonable time-distance relation (i.e., arrival times do not systematically increase with epicentral distance; Figure 9b). In contrast, the locations of the Pickblue-based hypoDD catalog can match the time-distance relation very well (Figure 9a), suggesting that these off-fault and extremely shallow earthquakes are likely mislocated in the Kuna catalog.

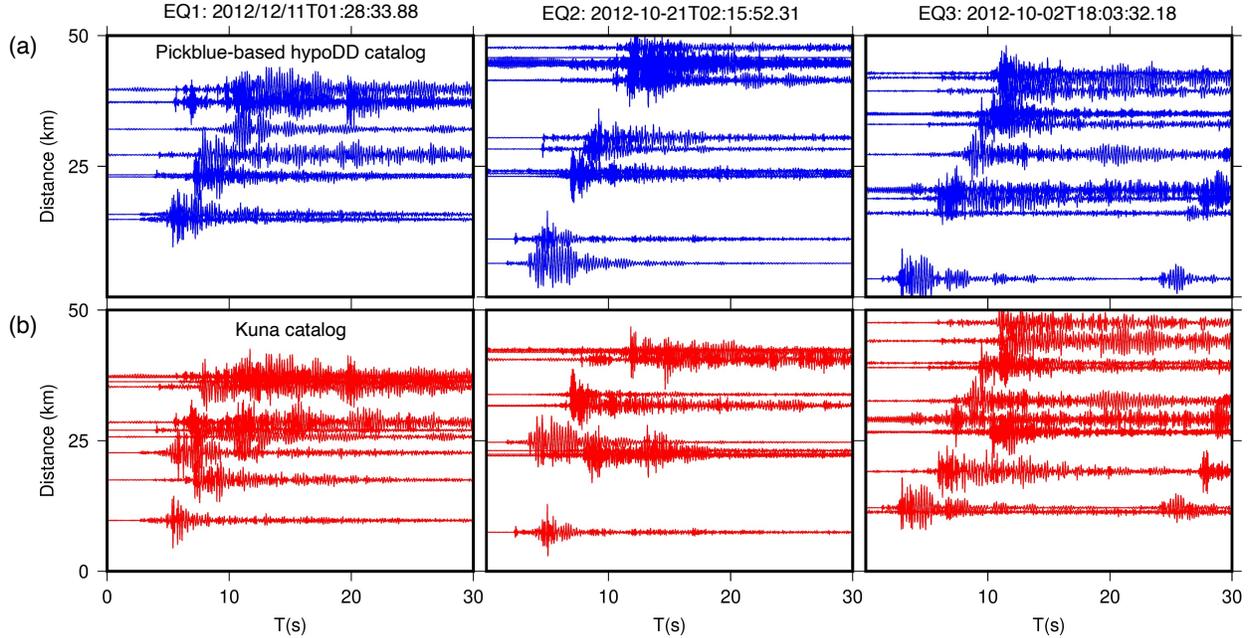

**Figure 9**. Investigation of the location reliability for three common events (EQs 1-3) with relatively large horizontal separations in the western BTF. (a) 4-15 Hz filtered vertical-component seismograms of the three events with locations allocated by the Pickblue-based hypoDD catalog at close stations (epicentral distance < 50 km). (b) Similar to (a), but the locations are allocated by the Kuna catalog.

We further check the common events with relatively large separations (> 4 km) in BR, where the Kuna catalog also includes extremely shallow events (Figure 10). We first examine the S-P time of a common earthquake (EQ4) that occurred ~2.2 km southeast of station BB140 (Figure 10). The S-P time of this event at station BB140 is ~1.7 s (Figure 11). By adopting the lower limit of P-wave velocity ($Vp_{low}$ = 4.20 km/s and $Vp_{low}/Vs_{low}$ = 2.1) that was used to build the Kuna catalog (Kuna, 2020), the estimated hypocenter distance to station BB140 is at least 6.5 km, coinciding with the hypocenter distance of ~6.7 km in the Pickblue-based hypoDD catalog. However, the depth of this event in the Kuna catalog is only ~0.3 km and the corresponding hypocenter distance is ~2.2 km, which is far less than the lower limit of 6.5 km. Besides, we also examine two other common events (EQ5 and EQ6) with a depth separation of ~12 km in the Kuna catalog (Figure 10). Based on the ~0.3 s difference in S-P time of the two events (Figure 11), we estimate the expected difference in hypocenter distance to station BB140 for the two events by adopting the lower and upper limits of P-wave velocity ($Vp_{low}$ = 4.20 km/s and $Vp_{low}/Vs_{low}$ = 2.1; $Vp_{upper}$ = 7.87 km/s and $Vp_{upper}/Vs_{upper}$ = 1.73) that was also used to build the Kuna catalog (Kuna, 2020). Results indicate that the two events' difference in hypocenter distance to station BB140 should be between 1.15 to 3.23 km, far less than the ~12 km depth separation of the two events in the Kuna catalog. In comparison, the Pickblue-based catalog provides more reasonable locations for the two events where their corresponding separation is ~1.36 km. Therefore, our analyses support that the extremely shallow earthquakes in the Kuna catalog are most likely mislocated events.

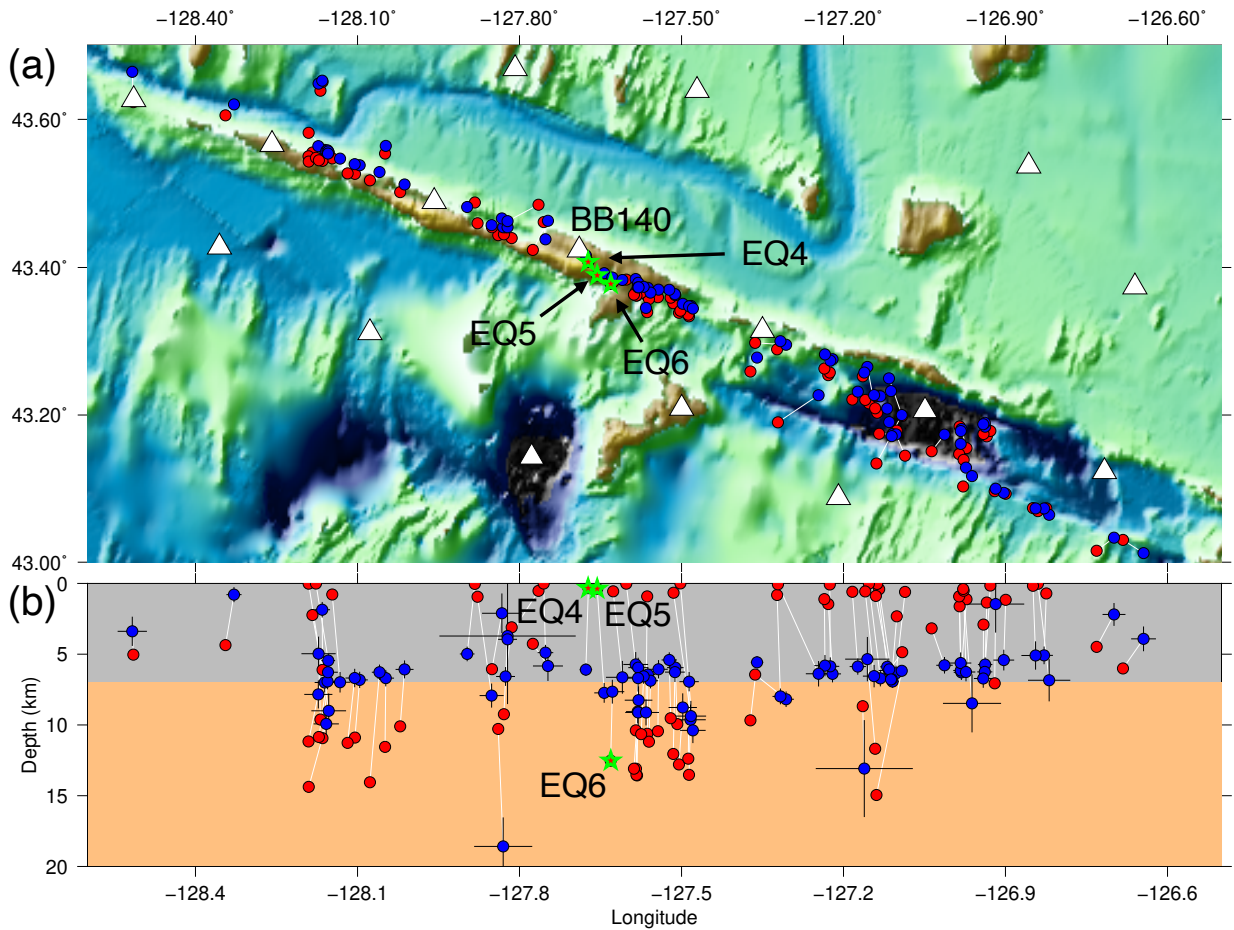

**Figure 10**. Comparison of the common events with relatively large separations between the Pickblue-based hypoDD catalog (blue dots) and the Kuna catalog (red dots) in BR. (a) Map view of the common events, which are linked by white lines. White triangles mark the OBS stations deployed by the Blanco experiment. Three green stars represent three common events (EQs 4-6), and their locations are further examined by inspecting their S-P times in Figure 11. (b) Projection

of these common events along the west-east direction. Black crosses indicate the relative errors of these events in the Pickblue-based hypoDD catalog.

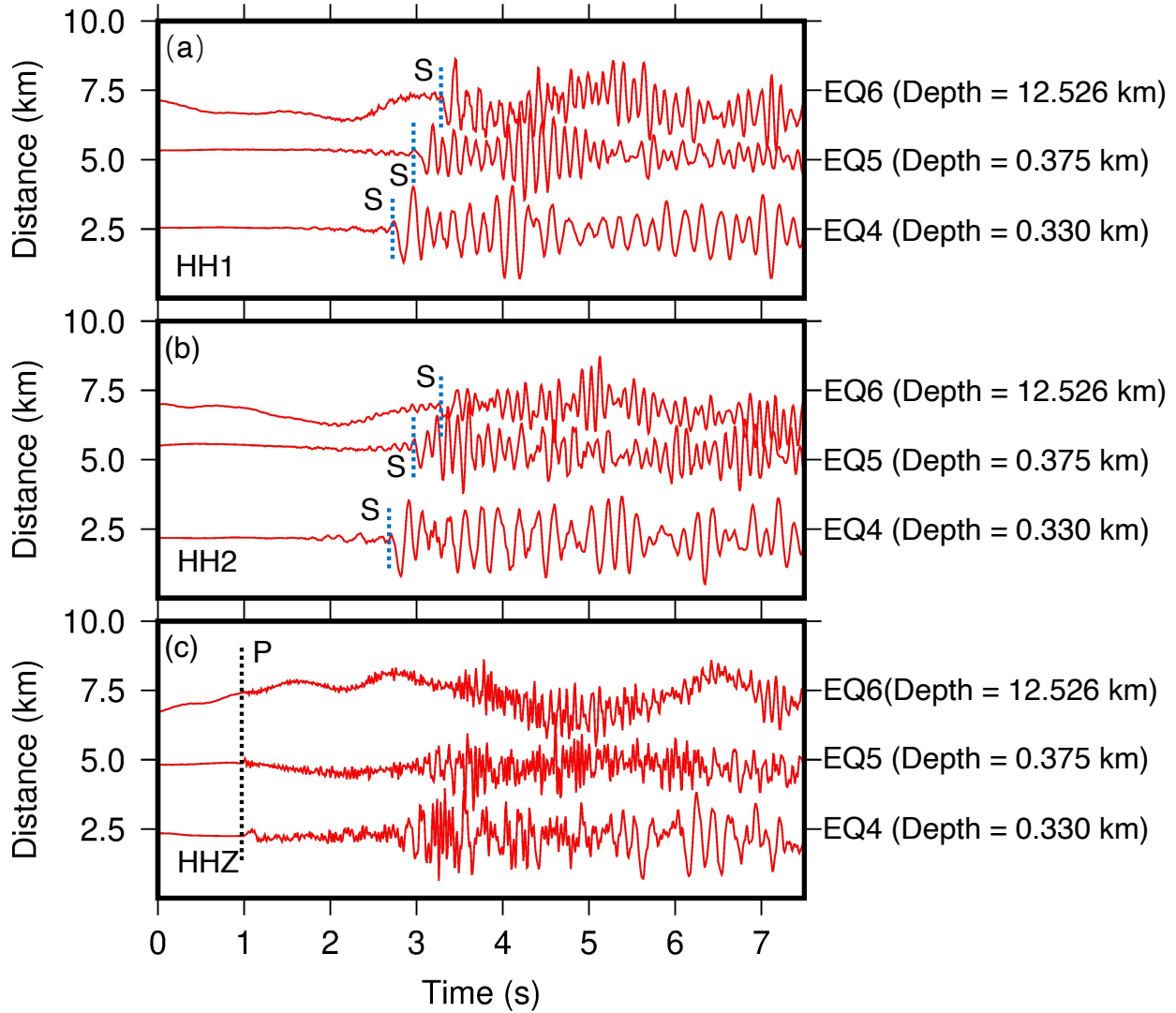

Figure 11. Inspection of the S-P times for three common events (EQs 4-6; 2012-12-05T01:03:27.41, 2012-11-06T18:17:50.24 and 2013-06-05T06:10:58.95) in BR. (a-b) Raw horizontal-component seismograms of the three events at the nearest station BB140 aligned by P-wave arrival times. Three blue dashed lines indicate the S-wave arrivals of the three events, respectively. The depths of the three events allocated by the Kuna catalog are marked as well (Kuna,

2020). (c) Similar to (a-b), but for vertical components. Black lines indicate the P-wave arrivals of the three events.

It should be mentioned that the Kuna catalog exhibits a wider seismicity gap around the Moho depth (~7 km) in BR than the Pickblue-based hypoDD catalog, especially below station BB140 where the width of the seismicity gap is up to ~5 km. Thus, we further analyze two events (EQs S1 and S2) below station BB140 in the Kuna catalog (Figure S4). The difference in the S-P times and depth separation of EQs S1 and S2 are ~0.3 s and ~5 km (Figures S4 and S5), respectively. Using the upper limit of P-wave velocity ($Vp_{upper}$ = 7.87 km/s and $Vp_{upper}/Vs_{upper}$ = 1.73) that was used to build the Kuna catalog (Kuna, 2020), the maximum difference of the hypocenter distance to station BB140 between the two events is estimated to be ~3.23 km, which is significantly less than the ~5 km depth separation in the Kuna catalog. Thus, our analysis indicates that earthquake depths in BR in the Kuna catalog are relatively poorly constrained and the separation between the crust and mantle earthquakes is overestimated, most likely because the Kuna catalog was built without utilizing S phases (Kuna, 2020).

The above analyses suggest that catalogs developed using different phase pickers have different advantages and shortcomings. The Kuna catalog, developed based on a traditional phase picker of STA/LTA (Kuna, 2020), provides relatively reliable hypocenters, although it documents a small portion of mislocated events (e.g., these extremely shallow earthquakes; Figure 4). However, the shortcoming of the traditional workflow is also obvious, since the Kuna catalog documents significantly fewer earthquakes (~7,000) than the two machine-learning-based hypoDD catalogs (~15,505 and ~26,944) (Figure 2c). Among the two machine-learning-based hypoDD catalogs, the OBSTransformer-based catalog documents ~11,000 more earthquakes than the Pickblue-based catalog (Figure 2c). However, the very different patterns of seismicity distribution and these mislocated deep earthquakes in the OBSTransformer-based hypoDD catalog imply a relatively high portion of P and S phase misidentifications (Figures 4b and 7b). This issue has a profound effect on the geological interpretation of the results, especially for the BR seismicity where the mantle earthquakes are mislocated in the crust (Figures 4b and 7b). In comparison, the Pickblue-based hypoDD catalog is better than the other two catalogs in multiple aspects. On the one hand, the Pickblue-based hypoDD catalog documents significantly more earthquakes than the Kuna catalog (Figure 2c). On the other hand, the Pickblue-based hypoDD catalog provides more reliable locations than the OBSTransformer-based hypoDD catalog (Figure 7). Note that the Pickblue-based hypoDD catalog also documents some unusually deep earthquakes in the eastern WBD and CD, but they exhibit significantly larger errors (Figures 8 and S3). Thus, the location errors in the Pickblue-based hypoDD catalog can serve as a quality controlling factor to easily filter out poorly-constrained events. In contrast, the mislocated events (e.g., these unusually deep earthquakes around -127.95° in longitude) in the OBSTransformer-based hypoDD catalog cannot be easily filtered out due to their analogous errors as other events (e.g., Figure S6). The different performances of these machine-learning-based phase pickers imply that it is necessary to carefully assess the reliability of the corresponding catalogs to prevent possible misinterpretations, especially when applied to regions without training samples.

## 4.2 Quality control for machine-learning-based workflow

Before the final relative relocation, we conduct strict quality control in earthquake selection, which includes four parameters: 1) travel time residual < 0.6 s; 2) station gap < 110°; 3) phase number > 11 and 4) have nearby station constraint (i.e., the three nearest stations have available P and/or S phases). With the first parameter of quality control, only ~4% (~2,000) of earthquakes are removed from the initial event database (i.e., Pickblue-based hypoinverse catalog), whereas the latter three parameters all can result in at least 40% of earthquakes being removed from the initial event database. Therefore, we further explore how the latter three parameters affect the reliability of the final locations. To do so, we rebuild three Pickblue-based hypoDD catalogs with the quality control of each catalog only adopting two of the latter three parameters (Figure S7).

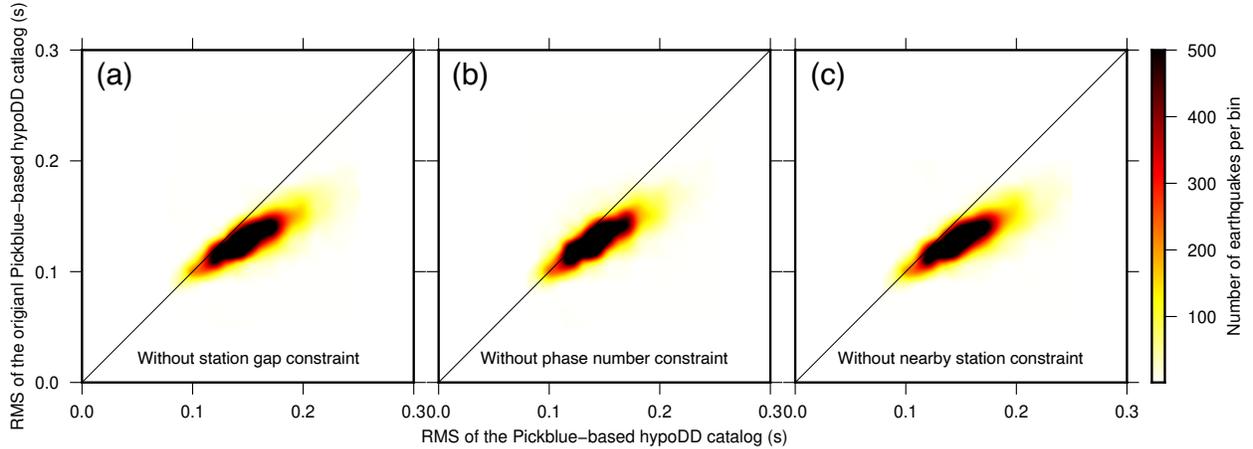

**Figure 12.** RMS comparison for common events between the original Pickblue-based hypoDD catalog and the other three Pickblue-based hypoDD catalogs without the constraints of station gap (a), phase number (b) and nearby station (c), respectively.

The two parameters of station gap and phase number are routinely used in quality control of machine-learning-based earthquake detection and location (Liu et al., 2020, 2023; Tan et al., 2021). The two newly developed Pickblue-based hypoDD catalogs without the constraints of the two parameters respectively document 17,642 and 20,154 earthquakes, which are ~2,000 and ~5,000 more events than the original Pickblue-based hypoDD catalog. However, the RMSs of most earthquakes listed in the two new hypoDD catalogs are higher than their common events in the original Pickblue-based hypoDD catalog (Figures 12a and 12b). We find that some newly included events in the new hypoDD catalog without the constraint of phase number fall in the shallow seismicity gaps (e.g., nearby station BS010; Figures S7 and S8) revealed by the original Pickblue-based hypoDD catalog. We compare the S-P times between two newly included events (EQs S3 and S4), which occurred ~7 km south of station BS010 (Figure S8). EQ S3 is located in the aforementioned seismicity gap with a depth of ~1.2 km, whereas EQ S4 is ~4.2 km deeper than EQ S3 (Figure S8). The difference in the S-P times of the two events at station BS010 is ~0.02 s (Figure S9). Using the upper limit of P-wave velocity ($Vp_{upper}$ = 7.87 km/s and $Vp_{upper}/Vs_{upper}$ = 1.73), the maximum difference of the hypocenter distance to station BS010 between the two events is estimated to be ~0.2 km. However, the actual difference in the hypocenter distance to station BS010 of the two events is ~1.6 km, implying that these newly included shallow earthquakes are likely mislocated. Therefore, our results further highlight the importance of the two frequently used constraints in building earthquake catalogs.

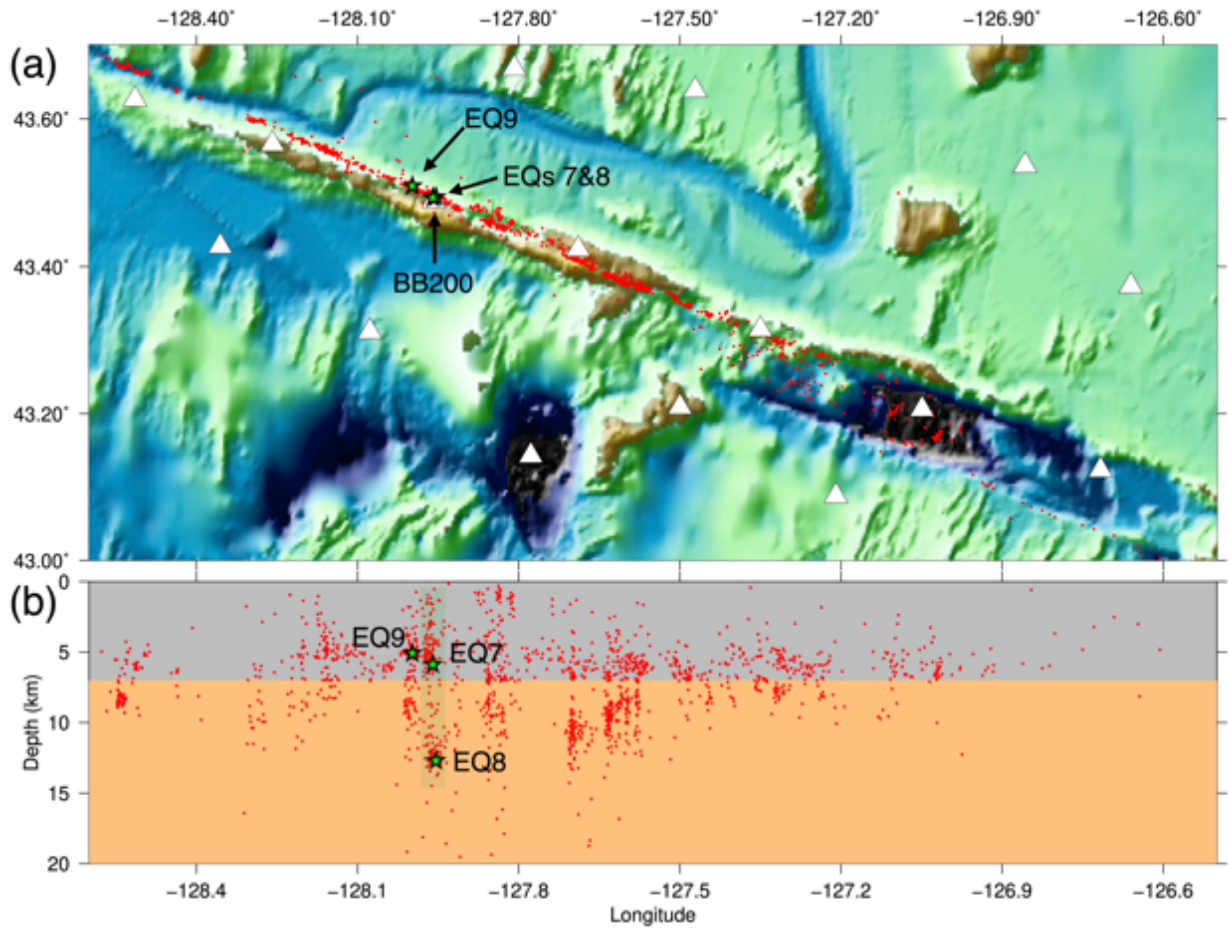

**Figure 13**. (a) Map view of the earthquakes (red dots) listed in the Pickblue-based catalog without nearby station constraints. White triangles mark the OBS stations deployed by the Blanco experiment. Three green stars represent epicenters of three newly included events (EQs 7-9), their locations are further examined in Figures 14 and 15. (b) Projection of these events exhibited in (a) along the west-east direction. Green shadow area marks the two earthquake clusters that are not observed in the original Pickblue-based hypoDD catalog.

Similarly, the majority of earthquakes listed in the new hypoDD catalog without nearby station constraints exhibit higher RMSs than the original Pickblue-based hypoDD catalog (Figure 12c). The new hypoDD catalog documents some deep (> 7 km) and distributed seismicity below SD (-

129.5° < longitude < -129°) and two earthquake clusters just below station BB200 with depths of ~5 km and ~12 km (Figure S4), which are not observed in the original Pickblue-based hypoDD catalog. Most newly included deep events (> 7 km) below SD exhibit relatively larger depth errors than shallow events (< 7 km; Figure S10), suggesting that these deep events are poorly located due to the lack of nearby station constraints. A similar issue is also seen in the original Pickblue-based hypoDD catalog, which documents some unusually deep and distributed earthquakes with significantly larger errors below the eastern WBD (Figure 8b). However, we also find that some events from the two newly added clusters below BR have available phases at the nearby stations, but the phases at the nearest station BB200 are not adopted during the absolute location, resulting in their depth errors being up to 10-20 km in the Pickblue-based hypoinverse catalog. We randomly select two events (EQs 7 and 8) from the two clusters and check their time-distance relations of the P-waves (Figure 13). Clearly, the P-wave arrival times of the two events at relatively close stations (i.e., epicenter distance < 80 km) can be reasonably associated except for the nearest station BB200 with a very large P-wave travel time residual of 0.7 s (Figures 14a and 14b). But this issue disappears for events further away horizontally from station BB200 (e.g., EQ9; Figure 14c), indicating that the large travel time residuals at station BB200 of the two newly added clusters are not due to site effect. The exact cause of this discrepancy remains to be determined. Note that the two clusters are also not observed in the Kuna catalog, suggesting that the lack of the two clusters in the original Pickblue-based hypoDD catalog is not caused by the parameter setting in our workflow. We compare the S-P times of the two events at the nearest station BB200. Results indicate that the two events have almost identical S-P time of ~1.75 s (Figure 15), but they are separated in depth by ~7 km (Figure 13), further suggesting that the two newly included earthquake clusters below station BB200 are mislocated. It should be mentioned that this parameter requiring nearby station constraint is rarely used for quality control in previous machine-learning-based earthquake detection and location studies. Here, our study illuminates that although this parameter significantly reduces earthquake numbers, it ensures a high-quality earthquake catalog.

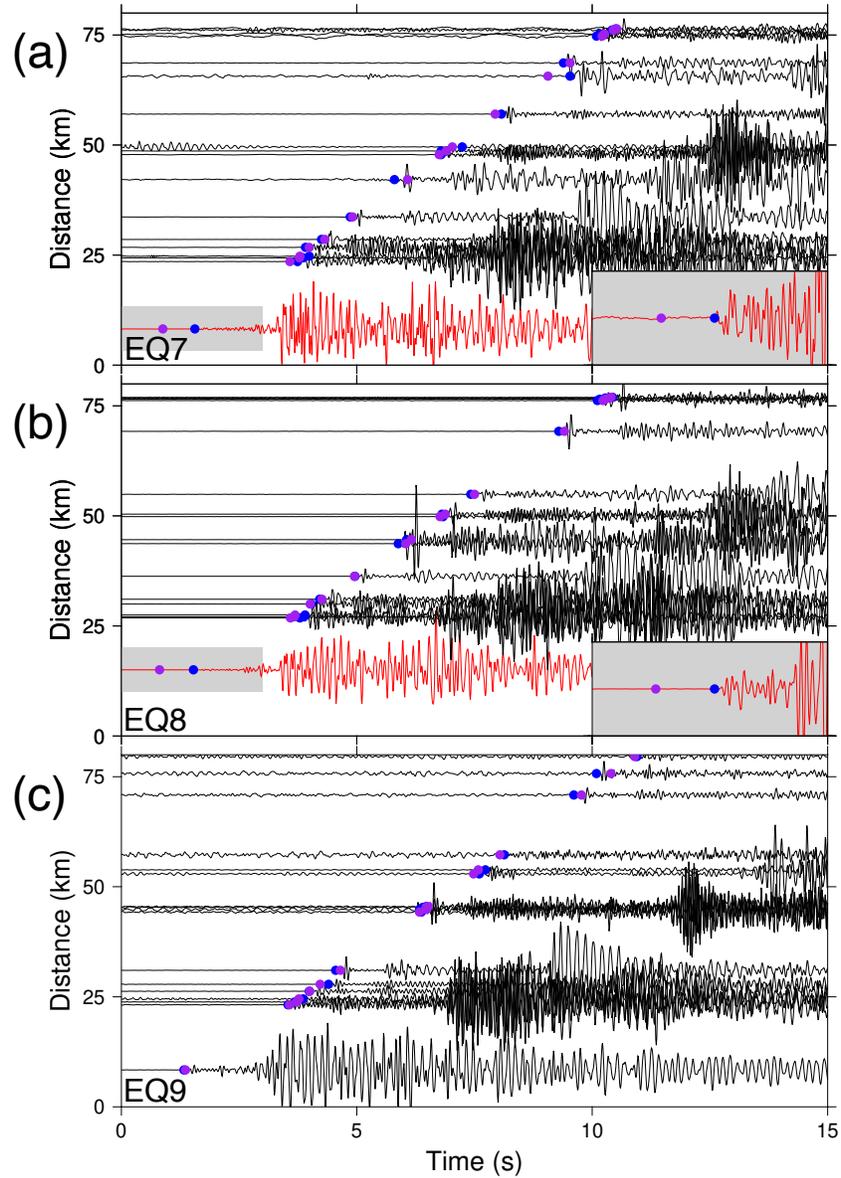

**Figure 14**. (a) 4-15 Hz filtered vertical-component seismograms of EQ7 (2013-05-24T16:27:02.36) with location allocated by the Pickblue-based hypoDD catalog without nearby station constraints (Figure 13). Blue and purple dots are picked and hypoinverse-estimated P-wave arrival times, respectively. Red seismogram is from the nearest station BB200 with a large P-wave arrival time residual. Gray shadow indicates the zoomed-in waveform window exhibited in the lower right corner. (b-c) Similar to (a), but for EQs 8 (2013-07-30T18:07:00.06) and 9 (2013-05-19T10:36:06.08).

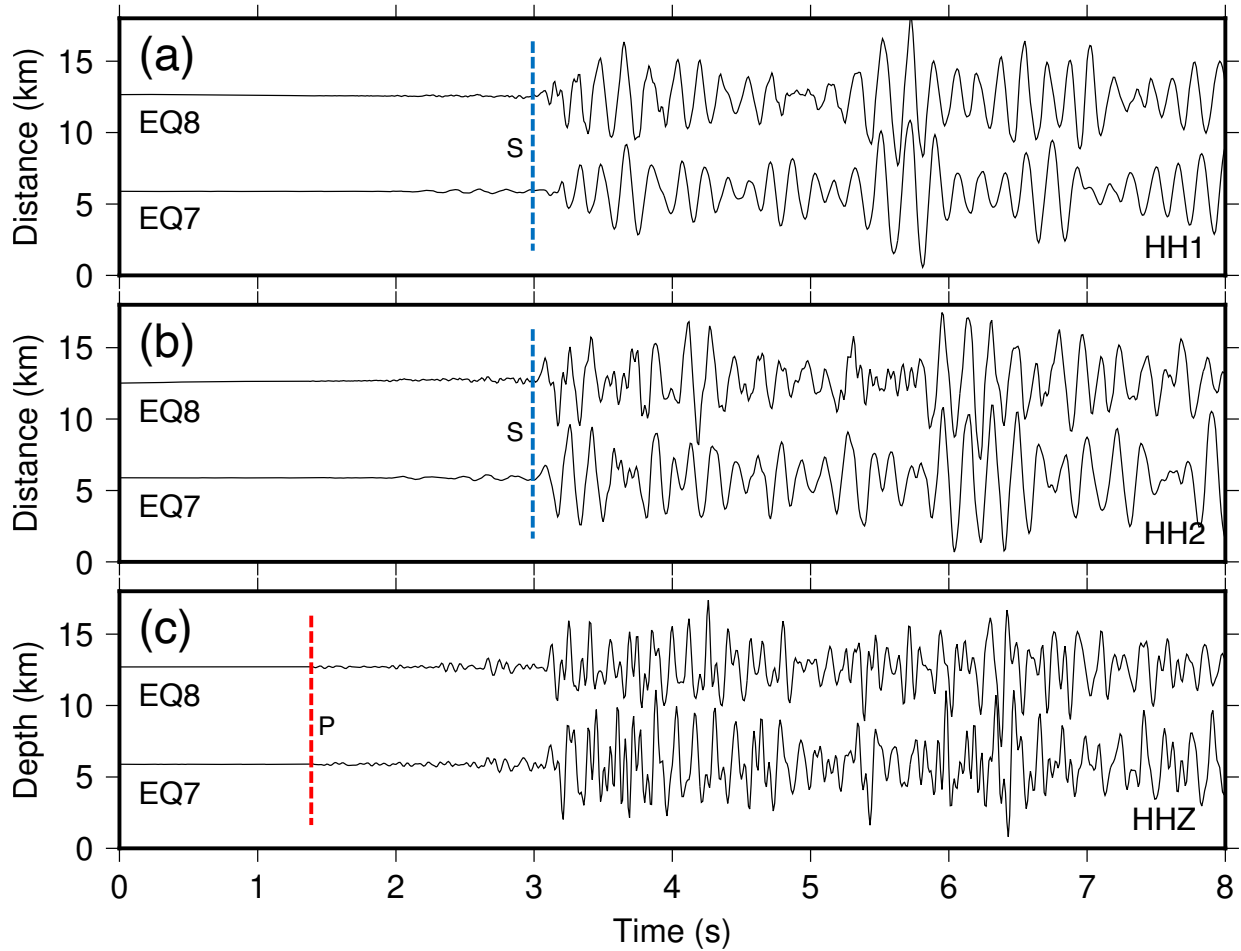

**Figure 15**. Inspection of the S-P times for EQs 7 (2013-05-24T16:27:02.36) and 8 (2013-07-30T18:07:00.06). (a-b) 4-15 Hz filtered horizontal-component seismograms of EQs 7 and 8 aligned by S-wave arrival times (blue dashed lines). (c) Similar to (a) and (b), but for vertical-component seismograms of EQs 7 and 8. Red dashed line marks the P-wave arrival time.

### 4.3 Seismicity distribution's implication for BTF's slip mode

Mislocated events may result in a wrong interpretation, especially for these unusually deep events. Thus, we exclude the poorly-constrained events from the Pickblue-based hypoDD catalog using a strict threshold before interpreting the seismicity distribution along BTF. Considering the mislocated earthquakes dominantly offset in depth due to the lack of nearby station constraints (Figure 8), we consider earthquakes with a depth error larger than the smallest depth error (0.036 km) plus standard deviation (0.086 km) as poorly-constrained events hence will not interpret them in this study. In total, 11,115 (71.7%) earthquakes are kept after this round of quality control, which is still ~40% more than the Kuna catalog. Compared with the original Pickblue-based hypoDD catalog, the general spatial patterns of these remaining events have no big change except the deep scatters and a seismicity cloud at a depth of ~9 km around -128.5° in longitude have been removed (Figure 16).

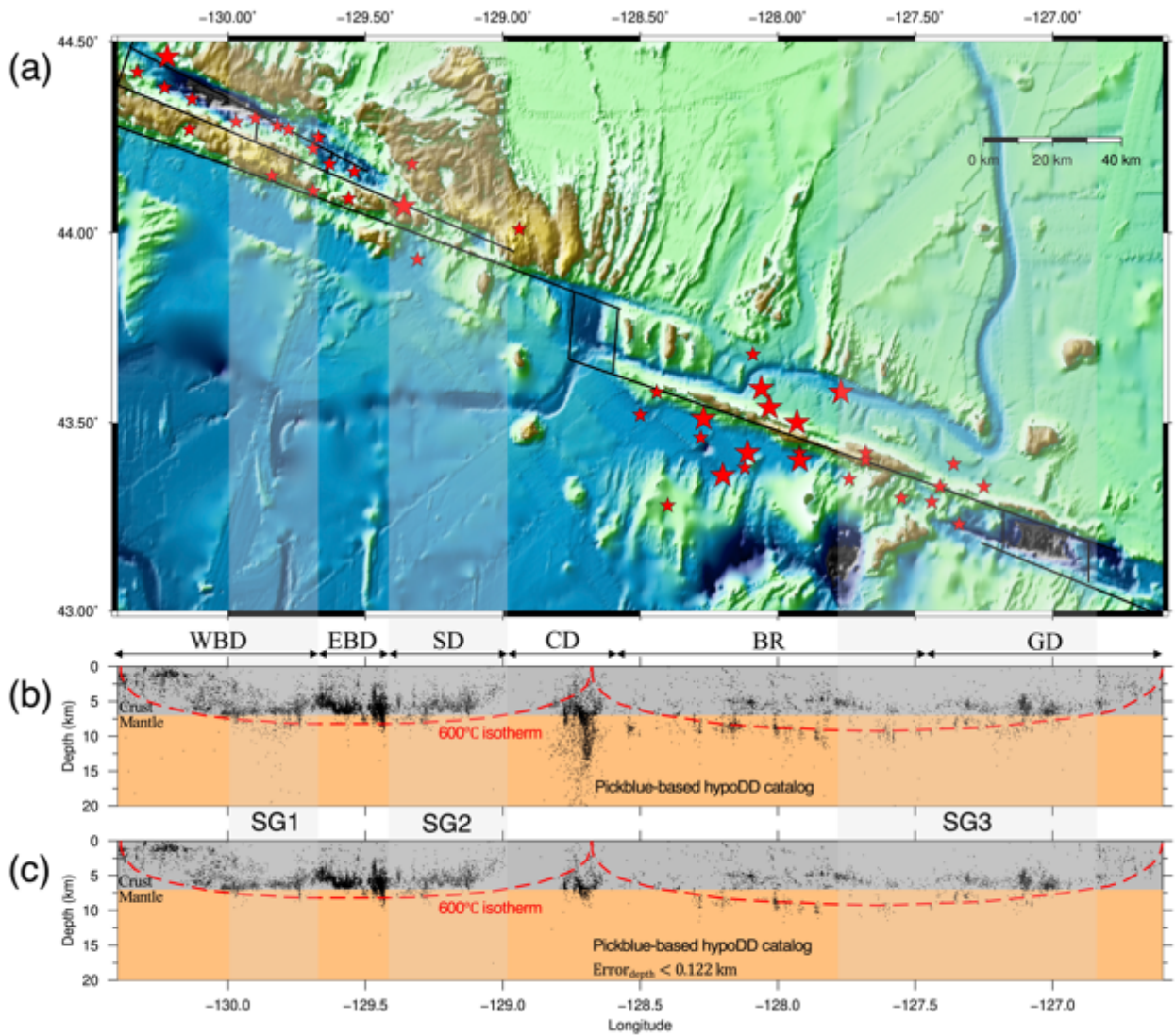

**Figure 16.** (a) Map view of our study area. Black lines outline the approximate locations of transform fault traces with their name being marked below the panel. Small and large red stars mark the 5.0 < Mw < 6.0 and Mw > 6.0 earthquakes documented in the CMT catalog, respectively. (b-c) Seismicity distribution along the east-west projection before and after removing earthquakes

with depth error > 0.122 km. Red dashed lines indicate the 600°C isotherm predicted by a simple half-space conductive cooling model. Gray shadow areas highlight the three major areas with seismicity gaps (SGs 1-3) at shallow depths.

The deficiency of seismic moment release is observed in most oceanic transform faults (Shi et al., 2022). In our study, the spatial distribution of earthquakes in the Pickblue-based hypoDD catalog reveals three major seismicity gaps (SGs 1-3) at shallow depths along BTF (Figure 16c), which may represent completely locked seismogenic zones during the interseismic period (Yu et al., 2021). However, the historical earthquakes since 1976 documented in the Global CMT catalog (red stars in Figure 16a) indicate that the corresponding fault segments of the three seismicity gaps did not host any Mw > 6.0 earthquakes except for the westernmost side of SG2 which hosted an Mw 6.3 earthquake in 2003. The fault segment of SG2 also exhibits significantly lower productivity of moderately large earthquakes (Mw > 5.5) than the other segments within BTF, suggesting that these shallow seismicity gaps are likely aseismic slip zones. The surface topography of the three seismicity gaps exhibits a strong extensional component, such as the widely developed depressions (Braunmiller & Nábělek, 2008), which are conducive to seawater infiltration into the upper crust. High pore pressure can effectively reduce the frictional contact area (Cappa et al., 2019), allowing aseismic slip at lower stress. This can limit the occurrence of large earthquakes, which is consistent with the deficiency of large earthquakes (Mw > 6.0) in the corresponding fault segments of the three seismicity gaps. However, these apparent shallow seismicity gaps may also be due to the limited one-year observational period. Therefore, further investigation (e.g., seismic tomography) is necessary to better understand their potentially different slip mode compared to other segments.

Numerous studies suggested that the seismogenic area accommodating brittle rupture is approximately bounded by the 600°C isotherm (Abercrombie & Ekström, 2001; W. Chen & Molnar, 1983; MCKENZIE et al., 2005). As a consequence, the maximum earthquake depth should exhibit a deepening trend along the oceanic transform faults from the ridge-transform intersection. However, some studies also suggested that the depth of earthquakes in oceanic transform faults can deepen to the 900°C isotherm due to seawater infiltration that results in serpentinization of mantle peridotite (Yu et al., 2021). In our study, earthquake depths show very different patterns between the western and eastern BTF (Figure 16c). Nearly all western earthquakes are located shallower than the 600°C isotherm predicted by a simple half-space conductive cooling model (Roland et al., 2010), whereas the eastern earthquakes, especially in BR, include multiple clusters that exceed the 600°C isotherm (Figure 16c). The CMT catalog indicates that BR accommodated many more Mw > 6.0 earthquakes (~80%) than other segments since 1976. Hence, these mantle earthquake clusters in BR can be explained by the serpentinization of mantle peridotite since the potentially deeper ruptures of these relatively large earthquakes can create conduits for seawater infiltration into the mantle. Therefore, our observations also suggest that the interaction between seawater infiltration and earthquake faulting may play an important role in controlling the slip mode of oceanic transform faults. It should be noted that the maximum depth of earthquakes below the central of BR in our catalog is ~10 km (Figure 16c), which can also be well explained by the thermal modelling incorporating hydrothermal cooling proposed by Roland et al. (2010) in which the depth limit of the 600°C isotherm is ~10 km. This highlights the necessity of further studies (e.g., seismic tomography) to confirm the mechanisms underlying these deeper earthquakes in BR. More importantly, the two alternative interpretations demonstrate the importance of careful examinations of earthquake depth constraint done in this study since mislocated earthquakes can result in very different conclusions. For instance, earthquakes in the Kuna catalog extended to ~13 km depth beneath BR, which is significantly deeper than the 600°C isotherm even for the thermal modelling incorporating hydrothermal cooling (Roland et al., 2010), hence would have incorrectly favored the seawater infiltration interpretation.

Finally, the middle BTF (i.e., CD) is relatively special compared with the western and eastern BTF since it has been considered as a small intra-transform spreading center underlain by partial melt (Adimah et al., 2024; Embley & Wilson, 1992) with a much shallower 600°C isotherm. We find that the depths of earthquakes in the middle BTF are far deeper than the 600°C isotherm, which may indicate that seawater infiltration cools the sparsely intruded magma (Kuna, 2020). A similar observation has previously been reported at the Lucky Strike segment in the slow-spreading Mid-Atlantic Ridge with earthquakes down to 14 km depth (Dusunur et al., 2009). Thus, these observations in our study highlight the varying effects of seawater infiltration within different segments along BTF.

5 Conclusions

In this study, we first adopt three different machine-learning-based phase pickers (i.e., EQTransformer, Pickblue and OBSTansformer) to build three hypoDD catalogs for the entire BTF based on a year-long OBS deployment. We then systematically compare the three machine-learning-based catalogs and a traditional workflow-based Kuna catalog. Results suggest that the Pickblue-based hypoDD catalog is the most ideal, as it documents more earthquakes and/or provides better-constrained earthquake locations than the other three catalogs. The varying performances of the three phase pickers suggest the importance of detailed assessment of catalogs developed by automatic workflows since mislocated earthquakes can result in very different conclusions.

The spatial distribution of earthquakes listed in the Pickblue-based hypoDD catalog reveals (1) Shallow seismicity gaps along some segments (e.g., WBD, SD, BR and GD) with significant extensional components which may reflect aseismic zones affected by seawater infiltration, and (2) Some deep earthquakes beyond the 600°C isotherm predicted by half-space conductive cooling beneath the BR, which can be explained by hydrothermal cooling or the serpentinization of mantle peridotite due to seawater infiltration along conduits created by potentially deeper ruptures of relatively large earthquakes.

Data Availability Statement

The OBS data used in this study are available at (Nábělek & Braunmiller, 2012). Historical earthquake data are provided by the Global Centroid Moment Tensor (GCMT) project available at https://www.globalcmt.org/. All figures were made in this study with Generic Mapping Tools (Wessel et al., 2013).

**Acknowledgement**

The authors thank Dr. Václav Kuna for providing the Kuna catalog (Kuna, 2020). This work was supported by the National Natural Science Foundation of China (42122060), the Hong Kong Research Grant Council Early Career Scheme (24305521), and the Croucher Tak Wah Mak Innovation Award.